\documentclass[preprintnumbers,
nofootinbib,
notitlepage,
 amsmath,amssymb,
]{revtex4-1}

\usepackage{graphicx}
\usepackage{dcolumn}
\usepackage{bm}

\makeatletter
\def\slash#1{{\mathpalette\c@ncel{#1}}} 
\makeatother

\newcommand\beq{\begin{eqnarray}}
\newcommand\eeq{\end{eqnarray}}

\newcommand\la{\langle}
\newcommand\ra{\rangle}

\begin{document}

\preprint{YITP-13-40}

\title{Double--spin asymmetry $A_{LT}$ in open charm production}

\author{Yoshitaka Hatta$^{\rm a}$}
\author{Koichi Kanazawa$^{\rm b,c}$}%
\author{Shinsuke Yoshida$^{\rm d,e}$}
\affiliation{\vspace{3mm}
  $^{\rm a}$Yukawa Institute for Theoretical Physics, Kyoto University, Kyoto 606-8502, Japan \\
  $^{\rm b}$Graduate School of Science and Technology, Niigata University, Ikarashi, Niigata 950-2181, Japan \\
  $^{\rm c}$Department of Physics, Barton Hall, Temple University, Philadelphia,
PA 19122, USA \\
$^{\rm d}$Theoretical Research Division, Nishina Center, RIKEN, Wako 351-0198, Japan \\
$^{\rm e}$Faculty  of Pure and Applied Sciences, University
of Tsukuba, Tsukuba, Ibaraki 305-8571, Japan
}%

\date{\today}

\begin{abstract}
\vspace{4mm}
In the collinear twist--three approach, we calculate for the first time the gluon contribution to double (longitudinal--transverse) spin asymmetry $A_{LT}$  for open charm production in proton--proton collisions  measurable at RHIC. Utilizing the Ward--Takahashi identity  for the non-pole part of the hard scattering amplitude, we derive a factorized, gauge invariant formula for the asymmetry. The result may be combined with the previous studies of single--spin asymmetry in the same channel and allows for a systematic analysis of three--gluon correlations inside a transversely polarized nucleon.

\end{abstract}

\pacs{Valid PACS appear here}
\maketitle



\section{Introduction}

The transverse--longitudinal, double--spin asymmetry $A_{LT}$ has long been known as a viable observable to probe the twist--three distributions in a  transversely polarized nucleon. Most of the theoretical studies so far have focused on the transverse quark structure function $g_T(x)=g_1(x)+g_2(x)$ \cite{Jaffe:1991kp,Tangerman:1994bb,Kanazawa:1998rw,Koike:2008du} accessible from inclusive measurement of $A_{LT}$ in polarized DIS and Drell--Yan experiments. As is well--known from the result of  the operator product expansion, $g_T(x)$ does not have a simple partonic interpretation, but is sensitive to  quark--gluon correlations inside the nucleon \cite{Shuryak:1981pi}. A dedicated analysis of the DIS data at JLab indeed revealed a significant content of such correlations \cite{Zheng:2004ce,Kramer:2005qe}.

If $A_{LT}$ is measured as a function of the transverse momentum $p_T$ of particles (jets) in  semi-inclusive processes, in the high--$p_T$ region it becomes sensitive  also to the `genuine twist--three' quark--gluon correlation functions. A variety of $p_T$--dependent processes have been calculated recently  \cite{Kotzinian:2006dw,Metz:2010xs,Kang:2011jw,Liang:2012rb,Metz:2012fq} including  photon $p^\uparrow p^\rightarrow \to \gamma(p_T)X$ \cite{Liang:2012rb} and hadron  $p^\uparrow p^\rightarrow \to h(p_T)X$ \cite{Metz:2012fq} productions in proton--proton ($pp$) collisions which are measurable  at RHIC.
     In fact, in these processes it is not enough to consider only $g_T(x)$ and the quark--gluon correlators. $A_{LT}$  receives contributions also from the ${\mathcal G}_{3T}(x)$ structure function (defined below in Eq.~(\ref{1})), which is the gluonic counterpart of $g_T(x)$, and the twist--three, three--gluon correlation functions. Although these gluonic functions are \emph{a priori} not suppressed compared with the quark ones, so far their theoretical treatment has  been scarce: The three--gluon correlators in a transversely polarized nucleon have only been discussed in the context of single--spin asymmetry  \cite{Ji:1992eu,Schmidt:2005gv,Kang:2008ih,Beppu:2010qn,Koike:2011mb}, whereas ${\mathcal G}_{3T}(x)$ has been discussed only in double transverse--spin asymmetry $A_{TT}$ in the Wandzura--Wilczek approximation \cite{Soffer:1997zy}. The detailed twist structure of ${\mathcal G}_{3T}$ as well as its relevance to the transverse--spin decomposition have  been recently elucidated in \cite{Hatta:2012jm}.

      In this paper we undertake the first calculation of the gluon contribution to $A_{LT}$ in the collinear factorization framework. Specifically, we shall compute, in the Feynman gauge, the numerator of $A_{LT}$ for open charm ($D$--meson) production in $pp$ collisions $p^{\uparrow}p^{\rightarrow}\to D(p_T)X$. This is an ideal channel to probe gluons because the   potential quark contribution from the subprocess $q\bar{q}\to g \to c\bar{c}$ is expected to be negligibly small as in the case of  single--spin asymmetry \cite{Kang:2008ih}.\footnote{In any case, the quark contribution can be retrieved  by a slight modification (inclusion of the $c$--quark mass in the hard coefficients) of the formula  obtained in \cite{Metz:2012fq}.} Our result will thus be the leading contribution to $A_{LT}$ in this channel which, with a suitable modeling of the gluon correlators, can be confronted with  future experimental results.


\section{Double spin asymmetry  in $p^{\uparrow}p^{\rightarrow}\to DX$: Setup}

\begin{figure}[htbp]
\begin{minipage}{0.45\hsize}
  \begin{center}
   \includegraphics[width=50mm]{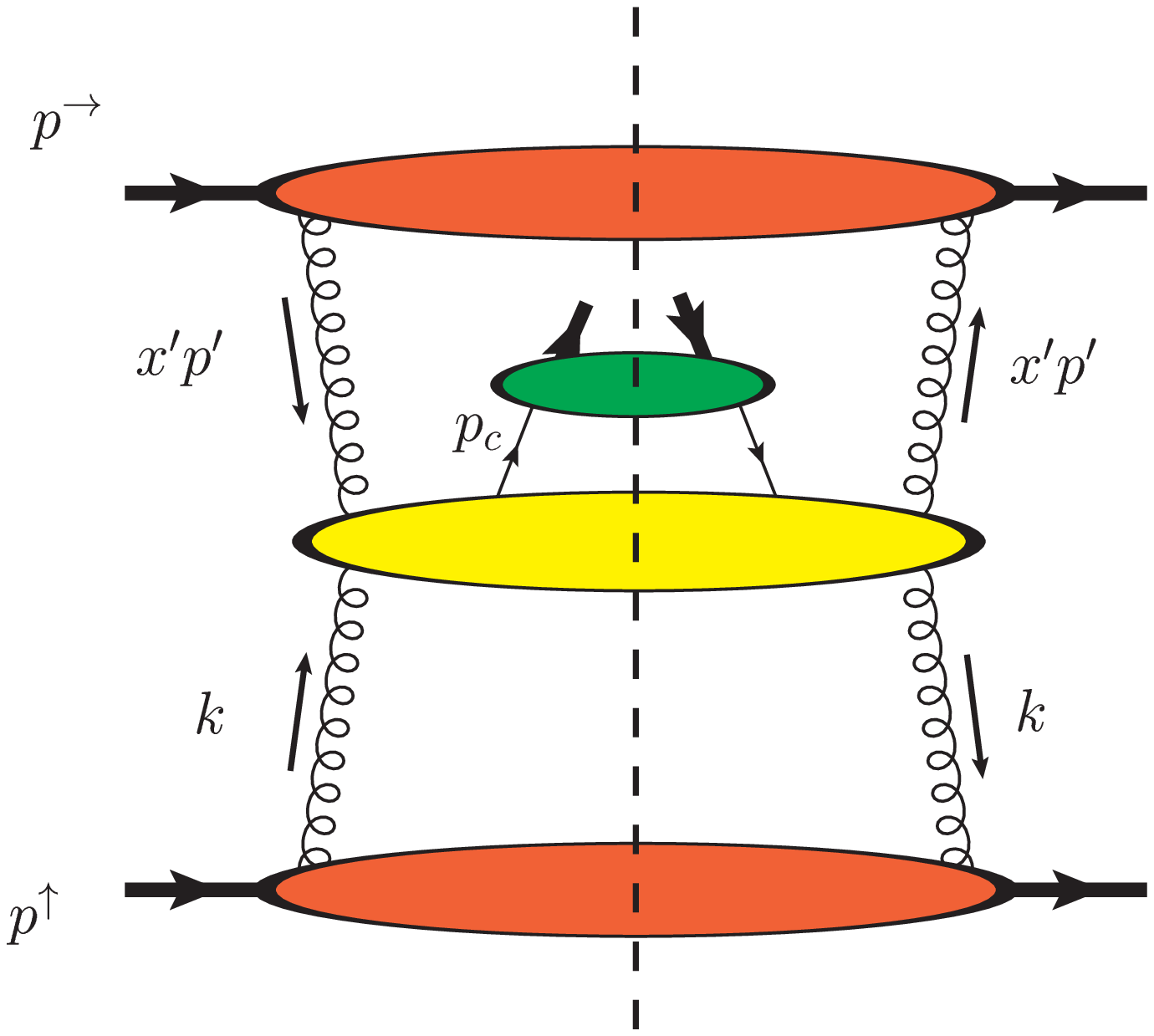}
  \end{center}
  \caption{Two--gluon contribution $W^1$ \label{a}}
 \end{minipage}
 \begin{minipage}{0.45\hsize}
  \begin{center}
   \includegraphics[width=50mm]{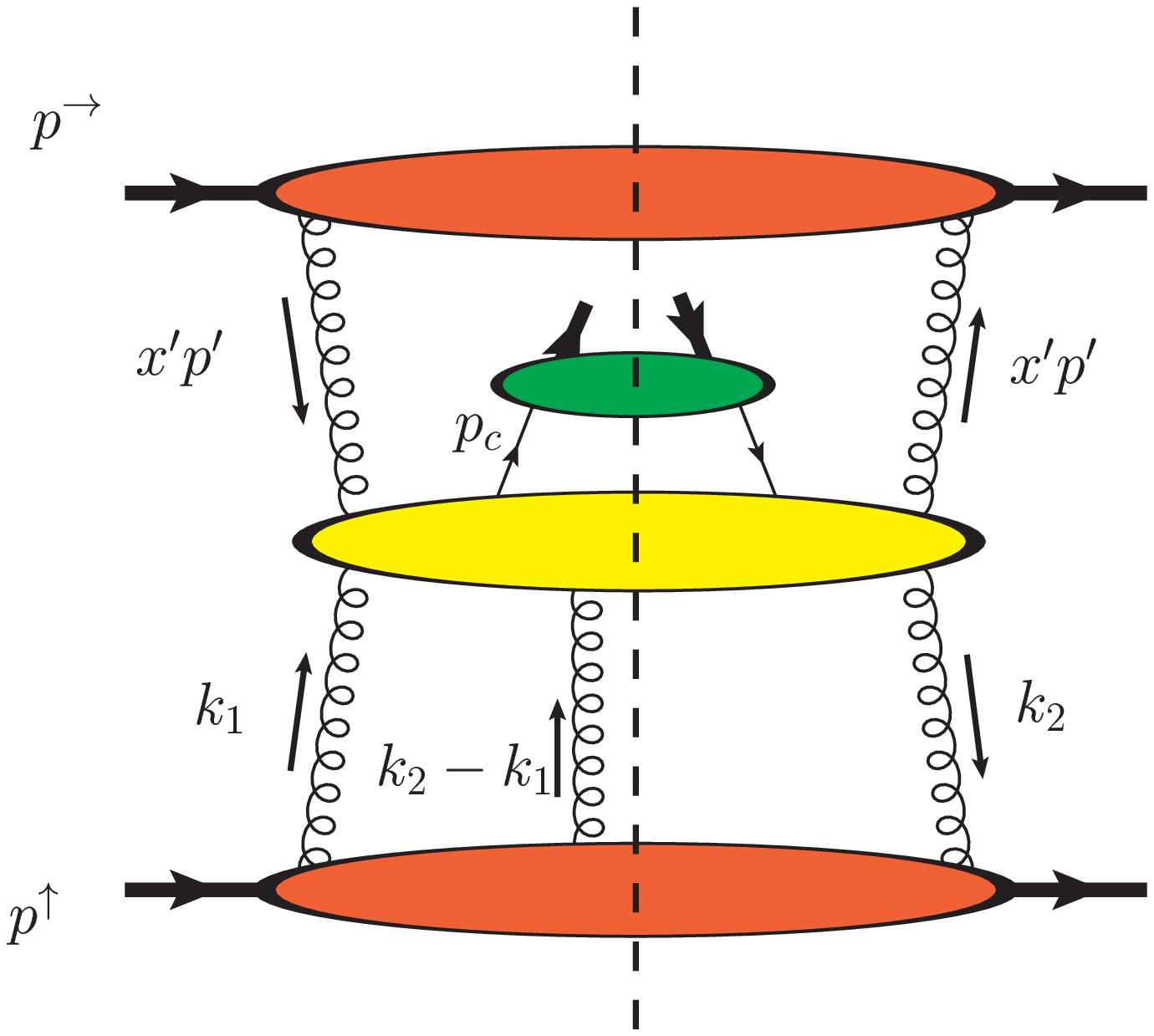}
  \end{center}
  \caption{Three--gluon contribution $W^2$ \label{b}}
 \end{minipage}
\end{figure}

The processes contributing to double--spin asymmetry (DSA) in $p^{\uparrow}p^{\rightarrow}\to DX$
 are graphically represented by Figs.~\ref{a}, \ref{b}.
The transversely polarized proton ($p^\uparrow$) is right--moving with momentum $p^\mu\approx \delta^\mu_+ p^+$, and the longitudinally polarized proton ($p^\rightarrow$) is left--moving with momentum $p'^\mu\approx \delta^\mu_- p'^-$.
The upper blobs represent the unpolarized $G(x)$ and polarized $\Delta G(x)$ gluon distributions
defined by the parametrization \cite{Ji:1992eu,Hatta:2012jm}\footnote{In \cite{Ji:1992eu}, ${\mathcal G}_{3T}(x)$ was denoted by $G_3(x)$.}
\beq
&&\int \frac{d\lambda}{2\pi} e^{i\lambda x} \langle PS| F^{n\alpha}(0) [0,\lambda n]F^{n\beta}(\lambda n) |PS\rangle
=-\frac{1}{2}xG(x)  (g^{\alpha\beta}-P^\alpha n^\beta -P^\beta n^\alpha)
 \nonumber \\
 && \qquad \qquad \qquad \qquad \qquad +\frac{i}{2}x\Delta G(x)  \epsilon^{nP\alpha\beta}n\cdot S +ix{\mathcal G}_{3T}(x) \epsilon^{n\alpha\beta \gamma}S_{\perp \gamma} + \cdots\,,  \label{1}
 \eeq
 where the notation $[0,z]$ represents the straight Wilson line from $0$ to $z^\mu$. The super--(sub--)scripts  $P$ and $n$ denote contraction with the momentum $P^\mu $ and the null vector  $n^\mu$ satisfying $P\cdot n=1$; e.g., $\epsilon^{np\alpha\beta}=\epsilon^{\mu\nu\alpha\beta}n_\mu P_\nu$ with the convention $\epsilon_{0123}=+1$. For the left--moving proton,
 $P^\mu =p'^\mu$ and $n^\mu=\delta^{\mu}_+/p'^-$.

 The middle blobs represent the hard scattering amplitude (cross section). In the case of single--spin asymmetry (SSA) $p^{\uparrow}p\to DX$, only the unpolarized gluon distribution of the left--mover is relevant. In order to generate a nonzero asymmetry, it is essential to extract the imaginary part of the amplitude by picking up the pole  of one of the internal propagators \cite{Qiu:1991wg}
 \beq
 \frac{1}{k^2+i\epsilon} = {\mathcal P}\frac{1}{k^2} -i\pi \delta (k^2)\,.
 \eeq
 Only the three--gluon amplitude (Fig.~\ref{b}) contains such an imaginary part, and this is why SSA is called a `genuine twist--three' observable.

 Of course, this standard mechanism of SSA also gives rise to a DSA of the same magnitude. However,  this does not concern us here because it has been already computed in \cite{Koike:2011mb}. The new feature which arises when the left--mover is longitudinally polarized is that the upper blob can be the polarized distribution $\Delta G(x)$. In this case, because of the relative factor of $i$ in the coefficient of $\Delta G(x)$, we must retain the {\it non}-pole part of the amplitude, which in particular means that both Fig.~\ref{a} and Fig.~\ref{b} are important.    We shall denote these non-pole parts as $S_{\mu\nu}^{ab}(k)$ and $S_{\mu\nu\lambda}^{abc}(k_1,k_2)$ ($a,b,..$ are color indices) respectively for the two figures.

  Finally the lower blobs represent the nonperturbative two-- and three--gluon  matrix elements
  \beq
M_{ab}^{\mu\nu}(k)&=&\int d^4\xi\, e^{ik\xi}\la pS | A_b^\nu(0)A_a^\mu(\xi) |pS\ra, \\
M_{abc}^{\mu\nu\lambda}(k_1,k_2)&=&g\int d^4\xi\int d^4\eta\,  e^{ik_1\xi+i(k_2-k_1)\eta}
\la pS | A_b^\nu(0) A_c^\lambda(\eta) A_a^\mu(\xi) |pS\ra\,,
\label{AAA}
\eeq
 taken in the transversely polarized proton state.
All in all,
the contributions to DSA from  Figs.~\ref{a}, \ref{b} can be written as the following convolution integrals
\beq
W^{\rm Fig.1} &=& \int \frac{dx'}{x'} \Delta G(x') \int \frac{dz}{z^2} D_c(z) \int \frac{d^4k}{(2\pi)^4} M_{ab}^{\mu\nu}(k) S_{\mu\nu}^{ab}
\left(k,x'p',P_h/z\right)\,, \label{fi} \\
W^{\rm Fig.2}&=& \frac{1}{2}\int \frac{dx'}{x'} \Delta G(x') \int \frac{dz}{z^2} D_c(z) \int \frac{d^4k_1}{(2\pi)^4} \frac{d^4k_2}{(2\pi)^4} M_{abc}^{\mu\nu\lambda}(k_1,k_2) S_{\mu\nu\lambda}^{abc}
\left(k_1,k_2,x'p',P_h/z\right)\,, \label{fig}
\eeq
where $D_c(z)$ is the $c$--quark fragmentation function into $D$--mesons. The symmetry factor $\frac{1}{2}$ in (\ref{fig}) accounts for the presence of two gluons on one side of the cut.

\section{Collinear expansion and Ward--Takahashi identity}

Our task now is to extract from (\ref{fig}) the twist--three part which contributes to DSA. We first note that the dominant region of the gluon momentum $k$ in Figs.~\ref{a}, \ref{b} is the region collinear to $p^\mu \approx \delta^\mu_{\ +}p^+$. We therefore decompose $k$ as
\beq
k^\mu=xp^\mu + \omega^{\mu}_{\ \sigma}k^\sigma\,, \qquad \omega^{\mu\nu}\equiv g^{\mu\nu}-p^\mu n^\nu\,,
\eeq
 where this time $n^\mu=\delta^\mu_{\ -}/p^+$ and $x \equiv k^+/p^+$.  We then perform the collinear expansion of the (non-pole) amplitudes up to ${\mathcal O}(\omega^3)$ which is the order needed for twist--three accuracy.

\beq
\hspace{-0.5cm}S_{\gamma\delta}(k) &=& S_{\gamma\delta}(xp)
+\omega^\mu_{\ \,\sigma}k^\sigma
\left.{\partial S_{\gamma\delta}(k) \over \partial k^\mu}\right|_{k=xp}
+{1\over 2}\omega^\mu_{\ \,\sigma}\,\omega^\nu_{\ \,\rho}k^{\sigma}k^\rho
\left.{\partial^2 S_{\gamma\delta}(k) \over
\partial k^\mu\partial k^\nu }\right|_{k=xp} \nonumber\\
&&+{1\over 6}\omega^\mu_{\ \,\sigma}
\,\omega^\nu_{\ \,\rho}\,\omega^\lambda_{\ \,\tau}k^{\sigma}k^{\rho}k^{\tau}
\left.{\partial^3 S_{\gamma\delta}(k) \over
\partial k^\mu \partial k^\nu \partial k^\lambda }\right|_{k=xp}+ \cdots,
\eeq

\beq
&& S_{\gamma\delta\kappa}(k_1,k_2) \nonumber \\
&& \quad =S_{\gamma\delta\kappa}(x_1p,x_2p)
+\omega^\mu_{\ \,\sigma}\left.\left(
k_1^{\sigma}{\partial  \over \partial k_1^\mu}
+k_2^\sigma{\partial \over \partial k_2^\mu}
\right)S_{\gamma\delta\kappa}(k_1,k_2)\right|_{k_i=x_ip} \nonumber\\
&&\quad +\omega^\mu_{\ \,\sigma}\,\omega^\nu_{\ \,\rho}\left.\left({1\over 2}k_1^{\sigma}k_1^{\rho}
{\partial^2  \over
\partial k_1^\mu\partial k_1^\nu}
+k_1^{\sigma}k_2^{\rho}
{\partial^2 \over
\partial k_1^\mu\partial k_2^\nu}+{1\over 2}k_2^{\sigma}k_2^{\rho}{\partial^2  \over
\partial k_2^\mu\partial k_2^\nu} \right)S_{\gamma\delta\kappa}(k_1,k_2)\right|_{k_i=x_ip}  \label{collinear1} \\
&& \quad +\omega^\mu_{\ \,\sigma}\omega^\nu_{\ \,\rho}\omega^\lambda_{\ \,\tau}\left.\biggl({1\over 6}
k_1^{\sigma}k_1^{\rho}k_1^{\tau}{\partial^3  \over
\partial k_1^\mu \partial k_1^\nu \partial k_1^\lambda }+{1\over 2}k_1^{\sigma}k_1^{\rho}k_2^{\tau}
{\partial^3  \over
\partial k_1^\mu \partial k_1^\nu \partial k_2^\lambda }  + \bigl\{k_1\leftrightarrow k_2\bigr\}
\biggr)S_{\gamma\delta\kappa}(k_1,k_2)\right|_{k_i=x_ip} + \cdots \,. \nonumber
\eeq
[The color indices are  omitted for simplicity.]

The multiple $k$--derivatives of $S_{\mu\nu}$ and $S_{\mu\nu\lambda}$ are very hard to evaluate in practice. Fortunately, they can be reduced to lower order derivatives by virtue of the Ward--Takahashi identity (WTI). For the two--gluon amplitude, we trivially have

\beq
k^\mu S^{ab}_{\mu\nu}(k) = k^\nu S^{ab}_{\mu\nu}(k)=0\,. \label{wa}
\eeq
Differentiating (\ref{wa}) twice with respect to $k$, we find relations such as

\beq
\left.{\partial^2 S^{ab}_{p\nu}(k) \over
\partial k^\mu\partial k^\lambda }\right|_{k=xp}&=&-{1\over x}
\left({\partial\over \partial k^\lambda}S^{ab}_{\mu\nu}(k)\Bigr|_{k=xp}
+{\partial\over \partial k^\mu}S^{ab}_{\lambda\nu}(k)\Bigr|_{k=xp}\right)\,,
\nonumber \\
\frac{\partial^2 S_{pp}^{ab}(k)}{\partial k^\mu \partial k^\lambda}\Bigg|_{k=xp} &=&\frac{1}{x^2}
\Bigl(S^{ab}_{\mu\lambda}(xp) + S^{ab}_{\lambda \mu}(xp)\Bigr)\,.
\eeq
[Remember our convention $S^{ab}_{p\nu} \equiv p^\mu S^{ab}_{\mu\nu}$.]
 The situation is more complicated in the three--gluon case.
 $S_{\mu\nu\lambda}^{abc}$ is defined such that
graphs in which incoming collinear lines merge into a single line are omitted. Due to this irreducibility property, it satisfies
 the following WTIs
\beq
k_1^\mu S_{\mu\nu\lambda}^{abc}(k_1,k_2) &=& -if^{abc}S_{\lambda \nu}(k_2) +
 G_{\lambda \nu}^{cba}(k_2-k_1,k_2)\,, \label{jun}
\\
k_2^\nu S_{\mu\nu\lambda}^{abc}(k_1,k_2) &=& - if^{abc}S_{\mu\lambda}(k_1) + \bigl(G_{\lambda \mu }^{cab}(k_1-k_2,k_1)\bigr)^* \,,
\\
(k_2-k_1)^\lambda S_{\mu\nu\lambda}^{abc}(k_1,k_2) &=& if^{abc}\bigl(S_{\mu\nu}(k_2) -S_{\mu\nu}(k_1)\bigr) \nonumber \\
 && + G_{\mu\nu}^{abc}(k_1,k_2) +\bigl(G_{\nu\mu}^{bac}(k_2,k_1) \bigr)^* \,, \label{last}
\eeq
 where $S_{\mu\nu}(k)\equiv \frac{1}{N_c^2-1}\delta^{ab} S^{ab}_{\mu\nu}(k)$.

\begin{figure}[t]
 \begin{center}
  \includegraphics[width=130mm]{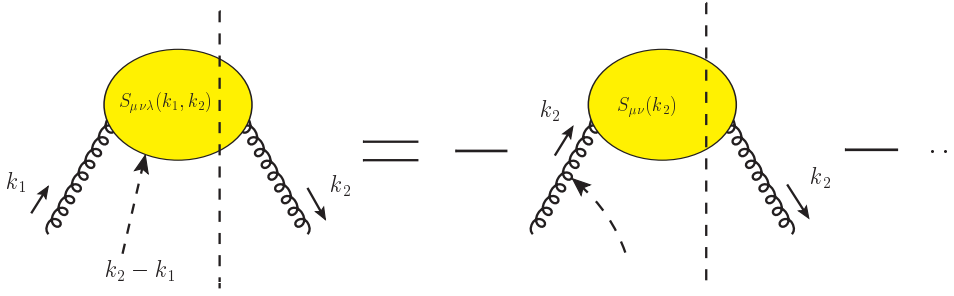}
 \end{center}
 \caption{Ward--Takahashi identity (\ref{last}) for the three--gluon amplitude. The dashed arrow
 represents an insertion of $\partial^\lambda A^c_\lambda$.  \label{wt}}
\end{figure}

 Comments are in order about the two--gluon  amplitude $S_{\mu\nu}$ on the right--hand--side. On one hand, the appearance of this term is easily understood diagramatically as shown in Fig.~\ref{wt}. On the other hand, it is actually a new element specific to the calculation of asymmetries which involve the {\it non}-pole part of the three--parton amplitude, as first observed in a related context \cite{Kanazawa:2013uia}. In the previous calculations of single--spin asymmetry which concerns the pole part, the three--parton amplitude satisfies a homogeneous WTI  without the two--parton amplitude  because of the on--shell condition for one of the internal propagators \cite{Eguchi:2006mc,Beppu:2010qn,Koike:2011mb}. We shall see that the inclusion of this term leads to a nontrivial gauge invariant contribution in the asymmetry (see, also, \cite{Kanazawa:2013uia}).

 The remaining terms, collectively denoted by $G$ in (\ref{jun})--(\ref{last}), come from the unphysical  polarizations of gluons whose detailed form  depends on the gauge choice.
 As we explain in Appendix A, the $G$--terms do not affect the result of this paper. We therefore neglect them and obtain, for instance, the relation
\beq
\left.{\partial S^{abc}_{\mu\nu p}(k_1,k_2) \over \partial k_1^\lambda}
\right|_{k_i=x_ip}={1\over x_2-x_1-i\epsilon}\left(S^{abc}_{\mu\nu\lambda}(x_1,x_2)
-if^{bca}{\partial\over \partial k_1^\lambda}S_{\mu\nu}(k_1)\Bigl.\Bigr|_{k_1=x_1p}\right)\,. 
\eeq
Other more complicated relations which we need are listed in Appendix B.

Making the most of these WTIs and performing a similar decomposition
\beq
M^{\mu\nu} &=& p^\mu p^\nu M^{nn} +\omega^{\mu}_{\ \gamma}p^\nu M^{\gamma n}+p^\mu \omega^{\nu}_{\ \gamma}M^{n\gamma} + \omega^{\mu}_{\ \gamma} \omega^{\nu}_{\ \delta} M^{\gamma\delta}\,, \nonumber \\
M^{\mu\nu\lambda}&=&p^\mu p^\nu p^\lambda M^{nnn} + \omega^{\mu}_{\ \gamma} p^\nu p^\lambda M^{\gamma nn} + \cdots\,,
\eeq
also for the soft matrix elements, we obtain after tedious calculations
\beq
W^{\rm Fig.1} + W^{\rm Fig.2}
&=&\int{dx'\over x'}\Delta G(x')\int{dz\over z^2}D_c(z)
\Biggl\{
\omega^\mu_{\ \,\sigma}\,\omega^\nu_{\ \,\rho}
\int {dx\over x^2}M_0^{\sigma\rho}(x)S_{\mu\nu}(xp) \nonumber \\
&&\qquad \quad -\frac{1}{2}\omega^{\mu}_{\ \,\sigma}\omega^{\nu}_{\ \,\rho}\omega^{\lambda}_{\ \,\tau}
\int{dx_1\over x_1}\int{dx_2\over x_2}M_{F\,abc}^{\sigma\rho\tau}(x_1,x_2)
{\mathcal P}{1\over x_2-x_1}S^{abc}_{\mu\nu\lambda}(x_1p,x_2p) \nonumber\\
&& \qquad \quad +i\omega^{\mu}_{\ \,\sigma}\omega^{\nu}_{\ \,\rho}\omega^{\lambda}_{\ \,\tau}\int{dx\over x^2}
\Bigl(M_{D}^{\sigma\rho\tau}(x)
-M_{\theta F}^{\sigma\rho\tau}(x)
\Bigr){\partial\over \partial k^{\lambda}}S_{\mu\nu}(k)\Bigl.\Bigr|_{k=xp} \Biggr\}\,,
\label{ted}
\eeq
where
\beq
M_0^{\sigma\rho}(x)&=& \int \frac{d\lambda}{2\pi} e^{i\lambda x} \la pS|F^{\rho n}(0) [0,\lambda n]F^{\sigma n}(\lambda n) |pS\ra\,,  \label{from} \\
M_{F\,abc}^{\sigma\rho\tau}(x_1,x_2)&=&-\int{d\lambda\over 2\pi}\int{d\mu\over 2\pi}
e^{i\lambda x_1+i\mu(x_2-x_1)}
\la pS |iF^{\rho n}_b(0)igF^{\tau n}_c(\mu n)iF^{\sigma n}_a(\lambda n)|pS\ra \,, \\
M_{D}^{\sigma\rho\tau}(x)&=&\int {d\lambda\over 2\pi} e^{i\lambda x}
\la pS |F^{\rho n}(0)[0,\lambda n]D^{\tau}(\lambda n)F^{\sigma n}(\lambda n)|pS\ra\,, \label{dtype} \\
M_{\theta F}^{\sigma\rho\tau}(x)&=&\int{d\lambda\over 2\pi}e^{i\lambda x}
\la pS |if^{bca}iF^{\rho n}_b(0)\Bigl(ig\int d\mu\theta(\mu-\lambda)F^{\tau n}_c(\mu n)\Bigr)
iF^{\sigma n}_a(\lambda n)|pS\ra \,. \label{to}
\eeq
In writing down (\ref{ted}), we actually only recovered the ${\mathcal O}(A^2)$ and ${\mathcal O}(gA^3)$ terms in the expansion of the matrix elements $\la FF\ra$, $\la gFFF\ra$ (including the expansion of the Wilson line). We then supplemented the result with the ${\mathcal O}(g^2A^4) \sim {\mathcal O}(g^4A^6)$ terms by hand which are not taken into account in Figs.~\ref{a}, \ref{b}, but which should organize themselves into  gauge invariant expressions.

\section{Spin--dependent cross section}
The matrix elements (\ref{from})--(\ref{to}) reduce to a set of invariant structure functions when projected out by the tensor $\omega^{\mu\nu}$. In the two--gluon part (first line of (\ref{ted}))   we find the function ${\mathcal G}_{3T}(x)$ defined in (\ref{1}) which, as already mentioned, is the gluonic counterpart of the $g_T(x)$ structure function.  As for the `F--type' correlator $\la FFF\ra$, we employ the following parametrization \cite{Beppu:2010qn}
\beq
\omega^\mu_{\ \ \sigma} \omega^\nu_{\ \ \rho} \omega^\lambda_{\ \ \tau}M_{F,abc}^{\sigma\rho\tau}(x_1,x_2) = \frac{N_c d_{abc}}{(N^2_c-1)(N^2_c-4)} O^{\mu\nu\lambda}(x_1,x_2) -\frac{if_{abc}}{N_c(N_c^2-1)} N^{\mu\nu\lambda}(x_1,x_2)\,,
\eeq
 where
 \beq
 &&O^{\mu\nu\lambda}(x_1,x_2) \nonumber \\
 &&= 2iM_N \left[O(x_1,x_2) g_\perp^{\mu\nu}\epsilon^{\lambda pnS_\perp}
 +O(x_2,x_2-x_1)g_\perp^{\nu\lambda}\epsilon^{\mu pn S_\perp} + O(x_1,x_1-x_2) g^{\mu \lambda}_\perp\epsilon^{\nu pn S_\perp} \right]\,,  \label{aaa}
 \eeq
 \beq
 && N^{\mu\nu\lambda}(x_1,x_2) \nonumber \\
 && = 2iM_N \left[N(x_1,x_2) g_\perp^{\mu\nu}\epsilon^{\lambda pnS_\perp}
 -N(x_2,x_2-x_1)g_\perp^{\nu\lambda}\epsilon^{\mu pn S_\perp} -N(x_1,x_1-x_2) g_\perp^{\mu\lambda}\epsilon^{\nu pn S_\perp} \right]\,. \label{bbb}
 \eeq
 [$M_N$ is the nucleon mass.]
 Due to the projection operator $\omega^\mu_{\ \ \sigma} \omega^\nu_{\ \ \rho} \omega^\lambda_{\ \ \tau}$, in (\ref{aaa}) and (\ref{bbb}) the indices $\mu,\nu,\lambda$ are practically restricted to the transverse ones ($\mu=1,2$, etc.), and accordingly we defined $g_\perp^{\mu\nu}=g^{\mu\nu}-p^\mu n^\nu -n^\mu p^\nu$.
 It is not necessary to introduce an independent parametrization of the `D--type' correlator $\la FDF\ra$ in (\ref{dtype}). In the last line of (\ref{ted}), one can write
 \beq
 \omega^\mu_{\ \ \sigma} \omega^\nu_{\ \ \rho} \omega^\lambda_{\ \ \tau} \bigl( M_D^{\sigma\rho\tau}(x)-M_{\theta F}^{\sigma\rho\tau}(x) \bigr) = -M_N \tilde{g}(x) \bigl(g_\perp^{\mu\lambda}\epsilon^{\nu pnS} -g_\perp^{\nu \lambda}\epsilon^{\mu pnS}\bigr)\,,
 \label{dif}
 \eeq
  where the function $\tilde{g}$ is defined in \cite{Hatta:2012jm}.\footnote{This should not be confused with the function $\tilde{g}$ (the same notation) defined in Refs.~\cite{Kang:2011jw,Liang:2012rb,Metz:2012fq}. There it is used  for the difference between D--type and F--type \emph{quark--gluon} correlators. Here (\ref{dif}) is the difference between D--type and F--type \emph{three--gluon} correlators. }
  Via the equation of motion, it can be fully expressed by the F--type correlator and other known distribution functions (see (3.2) and (3.22) of \cite{Hatta:2012jm}).

\begin{figure}[t]
 \begin{center}
  \includegraphics[width=120mm]{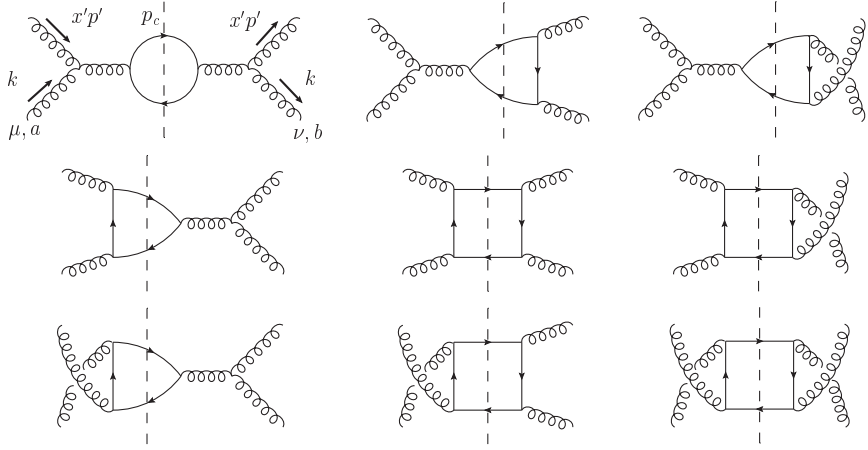}
 \end{center}
 \caption{Diagrams needed for the computation of $S^{ab}_{\mu\nu}(k)$. \label{s}}
\end{figure}

\begin{figure}[t]
 \begin{center}
  \includegraphics[width=120mm]{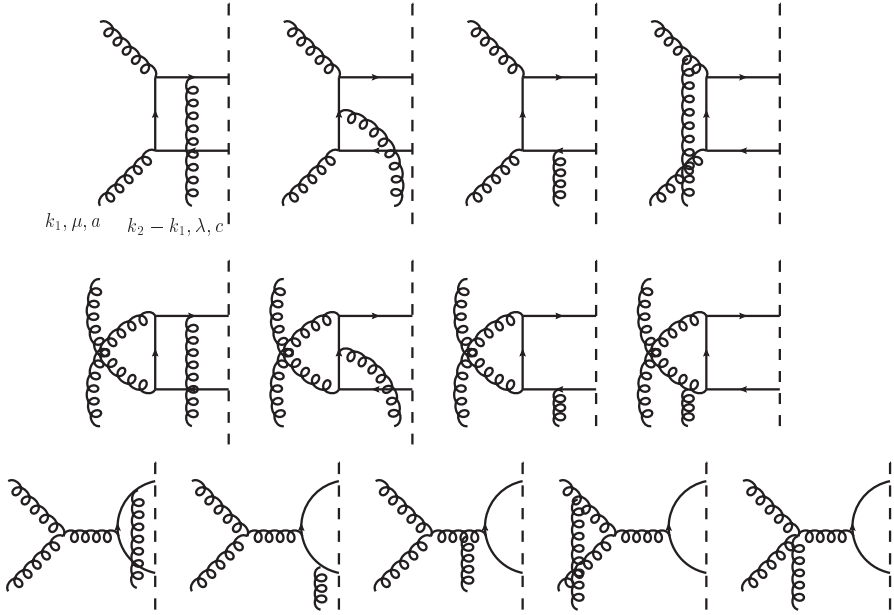}
 \end{center}
 \caption{Diagrams needed for the computation of $S^{abc}_{\mu\nu\lambda}(k_1,k_2)$. Only the left--hand--side of the cut is shown. \label{sl}}
\end{figure}

  Finally, we calculate the non-pole amplitudes $S_{\mu\nu}$, $dS_{\mu\nu}/dk^\lambda$ and $S_{\mu\nu\lambda}$ in the Feynman gauge to lowest order. The relevant diagrams are shown in Figs.~\ref{s} and \ref{sl}. Combining these results we arrive at the spin--dependent cross section which is the numerator\footnote{The spin--averaged cross section (the denominator of $A_{LT}$) may be found in \cite{Kang:2008ih,Koike:2011mb}. } of $A_{LT}$
\beq
P^0_h{d^3\Delta\sigma \over dP_h^3}
&=&{M_N\alpha^2_s\over s}(P_h\cdot S_{\perp})
\int{dx'\over x'}\Delta G(x')\int{dz\over z^3}D_c(z)\int dx\,
\delta(\tilde{s}+\tilde{t}+\tilde{u}) \nonumber\\
&&\times \Biggl[ {\mathcal G}_{3T}(x)\hat{\sigma}_{\rm 2} +
x \frac{d}{dx} \left(\frac{\tilde{g}(x)}{x^2}\right)
\hat{\sigma}_D+{\tilde{g}(x)\over x^2}\hat{\sigma}_D^\prime
 \nonumber\\
&& \qquad +\frac{1}{x}\int{dy\over y}{\mathcal P}{1\over x-y}\Biggl\{ O(x,y)\hat{\sigma}_{O}(x,y)
+O(x,x-y)\hat{\sigma}_{O}(x,x-y) \nonumber\\
&&\qquad \qquad +O(y,y-x)\bigl( \hat{\sigma}_{O}(x,-y)+\hat{\sigma}_{O}(x,y-x) \bigr)
\nonumber\\
&&\qquad \qquad \quad +N(x,y)\hat{\sigma}_{N}(x,y)+N(x,x-y)\hat{\sigma}_{N}(x,x-y) \nonumber \\ && \qquad \qquad \qquad
-N(y,y-x)\bigl(\hat{\sigma}_{N}(x,-y) -\hat{\sigma}_N(x,y-x)\bigr)
\Biggr\}
\Biggr]\,, \label{final}
\eeq
 where $s=(p+p')^2$.
 The various partonic cross sections are given by

\beq
\hat{\sigma}_{\rm 2}&=&{(\tilde{t}-\tilde{u})\over
\tilde{t}\tilde{u}}\Bigl({2\over N_c}+{1\over 2C_F}{\tilde{t}^2-4\tilde{t}\tilde{u}+\tilde{u}^2
\over \tilde{s}^2}\Bigr)\,,
\eeq
\beq
\hat{\sigma}_D&=&\Bigl({1\over N_c}{1\over \tilde{t}\tilde{u}^2}
-{1\over C_F}{1\over \tilde{s}^2\tilde{u}}
\Bigr)
{2(\tilde{t}\tilde{u}-2m_c^2\tilde{s})
(\tilde{t}^2+\tilde{u}^2)\over \tilde{t}\tilde{u}}\,,
\eeq
\beq
\hat{\sigma}_D^\prime&=&
{2\over N_c}{\bigl(\tilde{t}\tilde{u}(\tilde{t}^3-\tilde{u}^3)-2m_c^2\tilde{s}(\tilde{t}^3
-\tilde{t}\tilde{u}^2-2\tilde{u}^3)\bigr)
\over \tilde{t}^3\tilde{u}^3} \nonumber\\
&& \quad +{1\over 2C_F}{8m_c^2\tilde{s}(\tilde{t}^2-2\tilde{t}\tilde{u}-\tilde{u}^2)
-\tilde{t}(\tilde{t}-\tilde{u})(\tilde{t}^2+4\tilde{t}\tilde{u}+\tilde{u}^2)
\over \tilde{s}^2\tilde{t}^2\tilde{u}}\,,
\eeq
\beq
\hat{\sigma}_{O}(x,y)&=&{\tilde{s}(\tilde{t}^2+\tilde{u}^2)\bigl(\tilde{t}\tilde{u}(x+y)-2 m_c^2\tilde{s}x \bigr)
\over \tilde{t}^3\tilde{u}^3y}\Bigl({1\over N_c}-{1\over 2C_F}{3\tilde{t}\tilde{u}\over \tilde{s}^2}\Bigr)\,,
\eeq
\beq
\hat{\sigma}_{N}(x,y)&=&
{1\over 2N_c^2C_F}{\tilde{t}-\tilde{u}\over \tilde{t}^2\tilde{u}^2y}
\left[\Bigl((\tilde{t}^2+\tilde{u}^2)(x-y)-2\tilde{t}\tilde{u}y \Bigr)
-{2m_c^2\tilde{s}\over \tilde{t}\tilde{u}} \Bigl((\tilde{t}^2+\tilde{u}^2)(x-2y)-4\tilde{t}\tilde{u}y \Bigr)\right]
 \nonumber\\
&& \hspace{3mm} -{1\over 2C_F}{(\tilde{t}-\tilde{u})(\tilde{t}^2+\tilde{u}^2)\over \tilde{s}^2\tilde{t}^2\tilde{u}^2y}
\biggl[
\Bigl((\tilde{t}^2+\tilde{u}^2)(x-y)+\tilde{t}\tilde{u}(x-4y)\Bigr) \nonumber\\
&&\hspace{40mm}-{2m_c^2\tilde{s}\over \tilde{t}\tilde{u}}\Bigl((\tilde{t}^2+\tilde{u}^2)(x-2y)+\tilde{t}\tilde{u}(x-6y)\Bigr)
\biggr] \,,
\eeq
where
the Mandelstam variables
at the partonic level are defined as
\beq
\tilde{s}=(xp+x'p')^2\,, \hspace{5mm}\tilde{t}=(xp-p_c)^2-m^2_c\,,
\hspace{5mm}\tilde{u}=(x'p'-p_c)^2-m^2_c.
\eeq
\\

It is possible to eliminate both  ${\mathcal G}_{3T}(x)$ and $\tilde{g}(x)$ altogether from (\ref{final}) using the identities derived in \cite{Hatta:2012jm}. According to these identities,  ${\mathcal G}_{3T}(x)$ and $\tilde{g}(x)$ are decomposed into  the `Wandzura--Wilczek' part and the genuine twist--three part.\footnote{ See (3.15) and (3.22) of \cite{Hatta:2012jm}. The function $F(x,x')$ defined there is equal to $2N(x,x')$ here.}
 The  right--hand--side of (\ref{final}) can then be written by the polarized gluon distribution $\Delta G(x)$, the three--gluon correlation functions $N$ and $O$, and also the quark--gluon correlation function  $\sim \langle PS_\perp|\bar{q}\gamma^+F^{+\mu}q|PS_\perp\rangle$. The full expression does not appear to be particularly enlightening (though it may be useful in practice), so here we only note that in the `Wandzura--Wilczek approximation' which neglects all the three--parton correlators, the square brackets $[...]$ in (\ref{final}) reduces simply to
\beq
\frac{\hat{\sigma}_{\rm 2}+\hat{\sigma}_D^\prime }{2}\int^1_x \frac{dx'}{x'}\Delta G(x') -\frac{\hat{\sigma}_D}{2}\Delta G(x)\,.
\eeq
Namely, apart from the $c$--quark fragmentation function, the cross section is entirely expressed by the polarized gluon distribution.

\section{Conclusions}

In conclusion, we have calculated the gluon contribution to double--spin asymmetry $A_{LT}$ in open charm production measurable at RHIC. Due to the non-pole part of the hard amplitude involved in the calculation, it was crucial to properly handle the inhomogeneous Ward--Takahashi identity (\ref{jun})--(\ref{last}) (cf., \cite{Kanazawa:2013uia}) when evaluating the collinear expansion coefficients. The main result (\ref{final}) features, among others, the ${\mathcal G}_{3T}(x)$ structure function (see Eq.~(\ref{1})) which has hitherto received remarkably little attention  compared with its quark counterpart $g_T(x)$. This is actually the first time that a process in which ${\mathcal G}_{3T}(x)$ dominates the cross section has been identified. It therefore serves as the benchmark process for a systematic study of gluon correlations inside a transversely polarized nucleon. Note  that ${\mathcal G}_{3T}(x)$ carries nontrivial information of the gluon helicity distribution $\Delta G(x)$ in its Wandzura--Wilczek part \cite{Hatta:2012jm}.

If experiment should reveal a sizable asymmetry beyond that described by the WW part, $A_{LT}$ becomes a sensitive probe of the three--gluon correlation. In this sense, our work extends the previous calculation of the single--spin asymmetry in the same channel \cite{Kang:2008ih,Koike:2011mb}. Since the single and double spin asymmetries involve different linear combinations of the twist--three gluon correlation functions, a combined analysis of both experimental data can put  tighter constraints on the parametrization of these functions.  We hope that such an analysis is feasible at RHIC in future.

\begin{acknowledgments}
We thank Daniel Boer, Yuji Koike and Kazuhiro Tanaka for  discussions and comments.
 K.~K. is supported by JSPS Research Fellowships for Young Scientists (No.24.6959). S.~Y. is supported by
JSPS Strategic Young Researcher Overseas Visits Program
for Accelerating Brain Circulation (No.R2411).
\end{acknowledgments}

\appendix

\section{Ghost--like terms in the Ward--Takahashi identity}
In this Appendix we argue that the extra terms in the WTI (\ref{last}) can be neglected to the order of interest.
A straightforward calculation in the Feynman gauge shows that $G$ has the structure
\beq
G_{\mu\nu}^{abc}(k_1,k_2)=-(k_2-k_1)^\lambda
P_{\mu}^{\ \rho}(k_1) \Bigl( \epsilon^\perp_{\lambda \sigma} H_{\rho\nu}^{\sigma,abc}(k_1,k_2)
+\epsilon^\perp_{\rho \sigma} H_{\lambda\nu}^{\sigma,cba}(k_2-k_1,k_2) \Bigr)\,,
\eeq
where $P_{\mu\nu}(k)\equiv (k^2g_{\mu\nu}-k_{\mu}k_{\nu})$ and $\epsilon_{\alpha\beta}^\perp$ is the antisymmetric tensor in the transverse plane ($\alpha,\beta=1,2$).  $H$ is a certain function which satisfies a WTI $H_{\rho\nu}^{\sigma}(k_1,k_2) k_2^\nu=0$, but its explicit form is not needed for the present discussion.
Let us define
\beq
\tilde{S}_{\mu\nu\lambda}^{abc}(k_1,k_2)&\equiv& S_{\mu\nu\lambda}^{abc}(k_1,k_2)+V_{\mu\nu\lambda}^{abc}(k_1,k_2) \,,
\eeq
where
\beq
V_{\mu\nu\lambda}^{abc}(k_1,k_2)&=& P_{\lambda}^{\ \rho}(k_2-k_1) \Bigl( \epsilon_{\mu\sigma}^\perp H^{\sigma,cba}_{\rho\nu}(k_2-k_1,k_2)
+\epsilon^\perp_{\rho\sigma}H^{\sigma,abc}_{\mu\nu}(k_1,k_2)\Bigr)  \nonumber\\
&&+ P_{\lambda}^{\ \rho}(k_2-k_1) \Bigl( \epsilon_{\nu\sigma}^\perp H^{\sigma,cab}_{\rho\mu}(k_1-k_2,k_1)
+\epsilon_{\rho\sigma} H^{\sigma,bac}_{\nu\mu}(k_2,k_1)\Bigr)^*  \nonumber \\
 && +P_{\mu}^{\ \rho}(k_1) \Bigl( \epsilon^\perp_{\lambda \sigma} H_{\rho\nu}^{\sigma,abc}(k_1,k_2)
+\epsilon^\perp_{\rho \sigma} H_{\lambda\nu}^{\sigma,cba}(k_2-k_1,k_2) \Bigr)  \nonumber\\
&&+ P_{\nu}^{\ \rho}(k_2) \Bigl( \epsilon^\perp_{\lambda\sigma}H^{\sigma,bac}_{\rho\mu}(k_2,k_1)
+\epsilon^\perp_{\rho\sigma}H^{\sigma,cab}_{\lambda\mu}(k_1-k_2,k_1)\Bigr)^* \,.
\eeq
It can then be shown that
$\tilde{S}_{\mu\nu\lambda}$ satisfies WTIs without ghost--like terms
\beq
k_1^{\mu}\tilde{S}_{\mu\nu\lambda}^{abc}(k_1,k_2)&=&-if^{bca}S_{\lambda\nu}(k_2)\,, \nonumber\\
k_2^{\nu}\tilde{S}_{\mu\nu\lambda}^{abc}(k_1,k_2)&=&-if^{bca}S_{\mu\lambda}(k_1)\,, \nonumber \\
(k_2-k_1)^{\lambda}\tilde{S}_{\mu\nu\lambda}^{abc}(k_1,k_2)&=&if^{bca}\Bigl(S_{\mu\nu}(k_2)
-S_{\mu\nu}(k_1)\Bigr)\,.  \label{mod}
\eeq
Now consider the convolution in (\ref{fig})
\beq
&&\int {d^4k_1\over (2\pi)^4}\int {d^4k_2\over (2\pi)^4}
M_{abc}^{\mu\nu\lambda}(k_1,k_2)S^{abc}_{\mu\nu\lambda}(k_1,k_2) \nonumber\\
&&=\int {d^4k_1\over (2\pi)^4}\int {d^4k_2\over (2\pi)^4}
M_{abc}^{\mu\nu\lambda}(k_1,k_2)\bigl(\tilde{S}^{abc}_{\mu\nu\lambda}(k_1,k_2)
-V^{abc}_{\mu\nu\lambda}(k_1,k_2)\bigr)\,. \label{jk}
\eeq
Due to (\ref{mod}), the first term reproduces (\ref{ted}) with the replacement $S_{\mu\nu\lambda}^{abc} \to \tilde{S}_{\mu\nu\lambda}^{abc}$ which however can be replaced back to $S_{\mu\nu\lambda}^{abc}$ because the difference $V_{\mu\nu\lambda}(x_1p,x_2p)$, evaluated at on--shell, is proportional to either $p_\mu$, $p_\nu$ or $p_\lambda$, and it gives vanishing contribution (to twist--three accuracy) when  contracted with $M_F$. [Note that $p_\mu \omega^{\mu}_{\ \sigma}M_F^{\sigma...}=p_\sigma M_F^{\sigma...}$.] Concerning the second term of (\ref{jk}), the  product $M^{\mu\nu\lambda}V_{\mu\nu\lambda}$ contains a factor like $P_{\mu\rho}(k_1)A^\mu(\xi)$ which is equivalent to the insertion of $\partial_\mu (\partial^\mu A^\rho -\partial^\rho A^\mu)$. Via the equation of motion, this can be written as the sum of ${\mathcal O}(g)$ terms and a term featuring the longitudinally polarized gluon $\partial \cdot A$ (in the Feynman gauge). The former lead to ${\mathcal O}(g^2A^4)$ terms which are of the same order as those mentioned below (\ref{to}), and the latter can be shown not to contribute at twist--three.
 [Note that $H^\sigma_{\rho+}$ vanishes at on--shell due to the WTI $H^\sigma_{\rho \nu}k^\nu_2=0$.]
They are thus beyond the scope of this paper.

\section{Ward--Takahashi identities}

Here we list  WTIs which relate the derivatives of $S_{\mu\nu\lambda}$.
The $i\epsilon$--prescription in the denominators is suppressed for simplicity. It can be restored by the following substitutions: $x_1\to x_1-i\epsilon$, $x_2\to x_2+i\epsilon$, $x_2-x_1\to x_2-x_1-i\epsilon$.

\beq
&&\left.{\partial S^{abc}_{\mu p\lambda}(k_1,k_2) \over \partial k_1^\nu}
\right|_{k_i=x_ip}=-{1\over x_2}if^{abc}{\partial\over \partial k_1^\nu}
S_{\mu\lambda}(k_1)\Biggl.\Biggr|_{k_1=x_1p}\,, \hspace{5mm}
 \nonumber\\
&&\left.{\partial S^{abc}_{p\nu\lambda}(k_1,k_2) \over \partial k_1^\mu}
\right|_{k_i=x_ip}=-{1\over x_1}S^{abc}_{\mu\nu\lambda}(x_1,x_2)\,.
\eeq

\beq
&&\left.{\partial^2 S^{abc}_{\mu pp}(k_1,k_2) \over
\partial k_1^\nu\partial k_1^\lambda }\right|_{k_i=x_ip}
={1\over x_1}{1\over x_2}if^{bca}\Bigl({\partial\over \partial k_1^\lambda}S_{\mu\nu}(k_1)\Biggl.\Biggr|_{k_1=x_1p}
+{\partial\over \partial k_1^\nu}S_{\mu\lambda}(k_1)\Biggl.\Biggr|_{k_1=x_1p}\Bigr)\,, \nonumber\\
&&\left.{\partial^2 S^{abc}_{\mu pp}(k_1,k_2) \over
\partial k_1^\nu\partial k_2^\lambda }\right|_{k_i=x_ip}=-{1\over x_2}{1\over x_2-x_1}
\Bigl(S^{abc}_{\mu\lambda\nu}(x_1,x_2)
-if^{bca}{\partial\over \partial k_1^\nu}S_{\mu\lambda}(k_1)\Biggl.\Biggr|_{k_1=x_1p}\Bigr) \,.
\eeq
\beq
&&\left.{\partial^2 S^{abc}_{pp\lambda}(k_1,k_2) \over
\partial k_1^\mu\partial k_1^\nu}\right|_{k_i=x_ip}=
{1\over x_1}{1\over x_2}if^{abc}\left( \left.{\partial\over \partial k_1^\mu}S_{\nu\lambda}(k_1)\right|_{k_1=x_1p}
+\left.{\partial\over \partial k_1^\nu}S_{\mu\lambda}(k_1)\right|_{k_1=x_1p}\right)\,, \nonumber\\
&&\left.{\partial^2 S^{abc}_{pp\lambda}(k_1,k_2) \over
\partial k_1^\mu\partial k_2^\nu}\right|_{k_i=x_ip}={1\over x_1}{1\over x_2}
S^{abc}_{\mu\nu\lambda}(x_1,x_2)\,.
\eeq
\beq
&&\left.{\partial^2 S^{abc}_{p\nu p}(k_1,k_2) \over
\partial k_1^\mu\partial k_1^\lambda}\right|_{k_i=x_ip}=-{1\over x_1}{1\over x_2-x_1}
\Biggl(S^{abc}_{\mu\nu\lambda}(x_1,x_2)+S^{abc}_{\lambda\nu\mu}(x_1,x_2) \nonumber\\
&&\hspace{6cm}-if^{bca}{\partial\over \partial k_1^\mu}S_{\lambda\nu}(k_1)\Biggl.\Biggr|_{k_1=x_1p}
-if^{bca}{\partial\over \partial k_1^\lambda}S_{\mu\nu}(k_1)\Biggl.\Biggr|_{k_1=x_1p}\Biggr)\,, \nonumber\\
&& \left.{\partial^2 S^{abc}_{p\nu p}(k_1,k_2) \over
\partial k_1^\mu\partial k_2^\lambda}\right|_{k_i=x_ip}=-{1\over x_1}{1\over x_2-x_1}
\left(-S^{abc}_{\mu\nu\lambda}(x_1,x_2)
+if^{abc}\left.{\partial\over \partial k_2^\lambda}S_{\mu\nu}(k_2)\right|_{k_2=x_2p}\right)\,.
\eeq

\beq
&&\left.{\partial^3 S^{abc}_{ppp}(k_1,k_2) \over
\partial k_1^\mu \partial k_1^\nu \partial k_1^\lambda }\right|_{k_i=x_ip}=-{1\over x^2_1}{1\over x_2}
if^{bca}\left(\left.{\partial\over \partial k_1^\lambda}S_{\mu\nu}(k_1)\right|_{k_1=x_1p}
+\bigl(\mbox{5 permutations}\bigr)\right)\,, \nonumber\\
&&\left.{\partial^3 S^{abc}_{ppp}(k_1,k_2) \over
\partial k_1^\mu \partial k_1^\nu \partial k_2^\lambda }\right|_{k_i=x_ip}={1\over x_1}{1\over x_2}{1\over x_2-x_1}
\Biggl(S^{abc}_{\mu\lambda\nu}(x_1,x_2)+S^{abc}_{\nu\lambda\mu}(x_1,x_2) \nonumber\\
&&\hspace{6cm}-if^{abc}{\partial\over \partial k_1^\mu}S_{\nu\lambda}(k_1)\Biggl.\Biggr|_{k_1=x_1p}
-if^{abc}{\partial\over \partial k_1^\nu}S_{\mu\lambda}(k_1)\Biggl.\Biggr|_{k_1=x_1p}\Biggr)\,, \nonumber\\
&&\left.{\partial^3 S^{abc}_{ppp}(k_1,k_2) \over
\partial k_1^\mu \partial k_2^\nu \partial k_2^\lambda }\right|_{k_i=x_ip}={1\over x_1}{1\over x_2}{1\over x_2-x_1}
\Biggl(-S^{abc}_{\mu\nu\lambda}(x_1,x_2)
-S^{abc}_{\mu\lambda\nu}(x_1,x_2) \nonumber\\
&&\hspace{6cm}+if^{abc}{\partial\over \partial k_2^\lambda}S_{\mu\nu}(k_2)\Biggl.\Biggr|_{k_2=x_2p}
+if^{abc}{\partial\over \partial k_2^\nu}S_{\mu\lambda}(k_2)\Biggl.\Biggr|_{k_2=x_2p}\Biggr) \,. 
\eeq


\begin{thebibliography}{99}

\bibitem{Jaffe:1991kp}
  R.~L.~Jaffe and X.~-D.~Ji,
  Phys.\ Rev.\ Lett.\  {\bf 67}, 552 (1991).

\bibitem{Tangerman:1994bb}
  R.~D.~Tangerman and P.~J.~Mulders,
  hep-ph/9408305.

\bibitem{Kanazawa:1998rw}
  Y.~Kanazawa, Y.~Koike and N.~Nishiyama,
  Phys.\ Lett.\ B {\bf 430}, 195 (1998)
  [hep-ph/9801341].


\bibitem{Koike:2008du}
  Y.~Koike, K.~Tanaka and S.~Yoshida,
  Phys.\ Lett.\ B {\bf 668}, 286 (2008)
  [arXiv:0805.2289 [hep-ph]].

\bibitem{Shuryak:1981pi}
  E.~V.~Shuryak and A.~I.~Vainshtein,
  Nucl.\ Phys.\ B {\bf 201}, 141 (1982).

\bibitem{Zheng:2004ce}
  X.~Zheng {\it et al.}  [Jefferson Lab Hall A Collaboration],
  Phys.\ Rev.\ C {\bf 70}, 065207 (2004)
  [nucl-ex/0405006].

\bibitem{Kramer:2005qe}
  K.~Kramer, D.~S.~Armstrong, T.~D.~Averett, W.~Bertozzi, S.~Binet, C.~Butuceanu, A.~Camsonne and G.~D.~Cates {\it et al.},
  Phys.\ Rev.\ Lett.\  {\bf 95}, 142002 (2005)
  [nucl-ex/0506005].

\bibitem{Kotzinian:2006dw}
  A.~Kotzinian, B.~Parsamyan and A.~Prokudin,
  Phys.\ Rev.\ D {\bf 73}, 114017 (2006)
  [hep-ph/0603194].

\bibitem{Metz:2010xs}
  A.~Metz and J.~Zhou,
  Phys.\ Lett.\ B {\bf 700}, 11 (2011)
  [arXiv:1006.3097 [hep-ph]].

\bibitem{Kang:2011jw}
  Z.~-B.~Kang, A.~Metz, J.~-W.~Qiu and J.~Zhou,
  Phys.\ Rev.\ D {\bf 84}, 034046 (2011)
  [arXiv:1106.3514 [hep-ph]].



\bibitem{Liang:2012rb}
  Z.~-T.~Liang, A.~Metz, D.~Pitonyak, A.~Schafer, Y.~-K.~Song and J.~Zhou,
  Phys.\ Lett.\ B {\bf 712}, 235 (2012)
  [arXiv:1203.3956 [hep-ph]].

\bibitem{Metz:2012fq}
  A.~Metz, D.~Pitonyak, A.~Schaefer and J.~Zhou,
  Phys.\ Rev.\ D {\bf 86}, 114020 (2012)
  [arXiv:1210.6555 [hep-ph]].

\bibitem{Ji:1992eu}
  X.~-D.~Ji,
  Phys.\ Lett.\ B {\bf 289}, 137 (1992).

\bibitem{Schmidt:2005gv}
  I.~Schmidt, J.~Soffer and J.~-J.~Yang,
  Phys.\ Lett.\ B {\bf 612}, 258 (2005)
  [hep-ph/0503127].

\bibitem{Kang:2008ih}
  Z.~-B.~Kang, J.~-W.~Qiu, W.~Vogelsang and F.~Yuan,
  Phys.\ Rev.\ D {\bf 78}, 114013 (2008)
  [arXiv:0810.3333 [hep-ph]].

\bibitem{Beppu:2010qn}
  H.~Beppu, Y.~Koike, K.~Tanaka and S.~Yoshida,
  Phys.\ Rev.\ D {\bf 82}, 054005 (2010)
  [arXiv:1007.2034 [hep-ph]].

\bibitem{Koike:2011mb}
  Y.~Koike and S.~Yoshida,
  Phys.\ Rev.\ D {\bf 84}, 014026 (2011)
  [arXiv:1104.3943 [hep-ph]].


\bibitem{Soffer:1997zy}
  J.~Soffer and O.~V.~Teryaev,
  Phys.\ Rev.\ D {\bf 56}, 1353 (1997)
  [hep-ph/9702352].

\bibitem{Hatta:2012jm}
  Y.~Hatta, K.~Tanaka and S.~Yoshida,
  JHEP {\bf 1302}, 003 (2013)
  [arXiv:1211.2918 [hep-ph]].

\bibitem{Qiu:1991wg}
  J.~-W.~Qiu and G.~F.~Sterman,
  Nucl.\ Phys.\ B {\bf 378}, 52 (1992).

\bibitem{Kanazawa:2013uia}
  K.~Kanazawa and Y.~Koike,
  arXiv:1309.1215 [hep-ph];  arXiv:1307.0023 [hep-ph].



\bibitem{Eguchi:2006mc}
  H.~Eguchi, Y.~Koike and K.~Tanaka,
  Nucl.\ Phys.\ B {\bf 763}, 198 (2007)
  [hep-ph/0610314].

\end{thebibliography}

\end{document}